\documentclass[apj,superscriptaddress,twocolumn,nopreprintnumbers,floatfix,
superscriptaddress]{aastex631} 

\usepackage{amssymb}
\usepackage{amsmath}
\usepackage{graphicx}
\usepackage{dcolumn}
\usepackage{color,units}
\usepackage{xspace}
\usepackage{mathtools}
\usepackage{physics}
\usepackage{tensor}
\usepackage{bm}
\usepackage{lipsum}
\usepackage{revsymb}
\usepackage[normalem]{ulem}

\newcommand{\Z}{\mathcal{Z}}
\newcommand{\M}{\mathcal{M}}
\newcommand{\V}{\mathcal{V}}
\renewcommand{\L}{\mathcal{L}}

\newcommand{\mcl}{{M_\textrm{cl}}}
\newcommand{\rh}{{r_\textrm{h}}}

\newcommand{\trh}{{\tau_\textrm{rh}}}
\newcommand{\ph}{{\rho_\textrm{h}}}

\newcommand{\SPA}{School of Physics and Astronomy, Monash University, Clayton VIC 3800, Australia}
\newcommand{\OzGravMonash}{OzGrav: The ARC Centre of Excellence for Gravitational Wave Discovery, Clayton VIC 3800, Australia}
\newcommand{\OzGravANU}{OzGrav-ANU, Centre for Gravitational Astrophysics, College of Science, The Australian National University, ACT 2601, Australia}

\begin{document}

\title{The imprint of superradiance on hierarchical black hole mergers}
\shorttitle{Superradiance and hierarchical black hole mergers}
\shortauthors{Payne et al.}

\author{Ethan Payne}
\affiliation{\OzGravANU}
\affiliation{\SPA}
\affiliation{\OzGravMonash}
\correspondingauthor{Ethan~Payne}
\email{ethan.payne@ligo.org}

\author{Ling Sun}
\affiliation{\OzGravANU}

\author{Kyle Kremer}
\affiliation{TAPIR, California Institute of Technology, Pasadena, CA 91125, USA}
\affiliation{The Observatories of the Carnegie Institution for Science, Pasadena, CA 91101, USA}

\author{Paul D. Lasky}
\author{Eric Thrane}
\affiliation{\SPA}
\affiliation{\OzGravMonash}

\date{\today}

\begin{abstract}
Ultralight bosons are a proposed solution to outstanding problems in cosmology and particle physics: they provide a dark-matter candidate while potentially explaining the strong charge-parity problem. 
If they exist, ultralight bosons can interact with black holes through the superradiant instability.
In this work we explore the consequences of this instability on the evolution of hierarchical black holes within dense stellar clusters. 
By reducing the spin of individual black holes, superradiance reduce the recoil velocity of merging binary black holes, which, in turn, increases the retention fraction of hierarchical merger remnants. 
We show that the existence of ultralight bosons with mass $ 2\times10^{-14}\lesssim \mu/\textrm{eV} \lesssim2\times10^{-13}$ would lead to an increased rate of hierarchical black hole mergers in nuclear star clusters.
An ultralight boson in this energy range would result in up to $\approx60\%$ more present-day nuclear star clusters supporting hierarchical growth.
The presence of an ultralight boson can also double the rate of intermediate mass black hole mergers to $\approx0.08$\,Gpc$^{-3}$\,yr$^{-1}$ in the local Universe. 
These results imply that a select range of ultralight boson mass can have far-reaching consequences for the population of black holes in dense stellar environments. 
Future studies into black hole cluster populations and the spin distribution of hierarchically formed black holes will test this scenario.
\end{abstract}

\section{Introduction}
A number of extensions to the Standard Model of particle physics propose the existence of a theoretical \textit{ultralight boson} with a mass between $\mu\sim 10^{-33}$ -- $10^{-10}$\,eV~\citep{Arvanitaki2010String, Arvanitaki2010Stringb, Ringwald2013UDM}. 
These particles can provide solutions to outstanding problems in cosmology and fundamental particle physics such as by being viable candidates for dark matter~\citep{Jaeckel2010LE, Hui2017DM, Hu2000FDM} or solving the strong charge-parity problem~\citep{Peccei1977CPa, Peccei1977CPb, Weinberg1978CP}.
While ultralight bosons are not expected to strongly interact with particles from the Standard Model~\citep{Dine1981inter, Shifman1979inter, Kim1979inter}, the weak equivalence principle requires them to gravitate in a similar manner to visible matter~\citep{Detweiler1980unstable}. 

If an ultralight bosonic wave exists in the vicinity of a spinning black hole, the wave can become gravitationally bound to the black hole. 
The bosonic field may be amplified by the extraction of rotational energy from the black hole through a process known as \textit{superradiance}~\citep{Klein1929Super, Dicke1954Super, Zeldovich1971generation, Zeldovich1972amp, Press1972Super, Bekenstein1973SuperA, Bekenstein1998SuperB, Brito2015Super, Brito2020superradiance}.
Superradiant amplification of the bosonic field can be unstable, leading to the exponential growth of the bosonic wave~\citep{Press1972Super}, forming a macroscopic quantum object, or boson \textit{cloud}~ \citep[e.g.,][]{Balakumar2020QSuper}. 
This \textit{superradiant instability} occurs most rapidly when the Compton wavelength of the particle is comparable to the outer horizon radius of a spinning black hole, and only ceases when the angular frequency of the black hole's rotation equals the frequency of the bosonic wave~\citep{Arvanitaki2010String, Bekenstein1973SuperA, Brito2015Super, Bekenstein1998SuperB}.
The final astrophysical system, following unstable black hole superradiance, is a lighter black hole with a reduced spin, surrounded by a macroscopic boson cloud~\citep{Brito2017search}. This matter configuration can also lead to long-lasting, nearly monochromatic gravitational-wave radiation from the rotation of the boson cloud-black hole system~\citep{Yoshino2013grav}, often referred to as a continuous gravitational-wave signal. 

Directed searches for continuous waves from ultralight boson clouds around black hole merger remnants have not yet been undertaken because the signal is likely too weak to observe with current observatories~\citep{Isi2018direct, Sun2020}.
Nonetheless, non-detections of an incoherent stochastic background from the continuous-wave emission have placed tentative constraints on boson masses of $(2$--$3.8)\times10^{-13}$\,eV given particular assumptions about the black hole formation spin distribution~\citep{Tsukada2019stochA, Brito2017stoch, Tsukada2021stoch}.

Furthermore, by studying the spins of the population of resolved binary black hole (BBH) mergers, constraints exclude boson masses between $(1.3$ -- $2.7)\times10^{-13}$\,eV assuming negligible boson self-interactions~\citep{Ng2019constraints, Ng2021constraints}.
Other observations of black holes such as the recent images of M87$^*$ from the Event Horizon Telescope~\citep{Akiyama2019cqa, Davoudiasl2019}, and radial velocity and photometric data from Cygnus X-1~\citep{Iorio2007at, Orosz2011, Reid2011, Middleton2016, Isi2018direct} have excluded the presence of ultralight bosons in different regions of the boson mass parameter space. 
Ultimately, there is no strong evidence yet for either the existence or absence of ultralight bosons.

A population of low-spin black holes produced by superradiance would have wide-ranging astrophysical consequences. 
In dense stellar environments, such as nuclear star clusters or globular clusters, low-spin, first-generation (1G) black holes can merge hierarchically to form second-generation (2G) black holes~\citep{Zwart2000Dense, Miller2009, Downing2010Dense, Rodriguez2015Dense, Rodriguez2016DenseB, Rodriguez2018DenseC, Antonini2016Nscs, Petrovich2017, Fragione2018, Antonini2018hierarchical, Arca-Sedda2018, Kremer2020b}. 
By decreasing black-hole spins through superradiance, the gravitational-wave recoil velocities are reduced~\citep{Campanelli2006kick, Campanelli2007kick, Campanelli2007kickB, Gonzalez2007kick, Gonzalez2007kickB, Lousto2010kick, Lousto2012kick, Varma2018kick}. 
Lower recoil velocities lead to a higher retention fraction of binary black hole merger remnants~\citep{Merritt2004conseq, Varma2020extractingKick}, and subsequent enhancement of hierarchical black hole growth as a result~\citep{Antonini2018hierarchical,Rodriguez2018DenseC,Rodriguez2019repeated}.

The remainder of the manuscript is structured as follows. In Sec.~\ref{sec:sr}, we discuss the theory of black hole superradiance in the context of scalar bosons and its impact on individual black holes.\footnote{We focus on scalar (spin-0) bosons~\citep{Damour1976, Ternov1978scalar, Detweiler1980unstable, Yoshino2013grav}.
However, the general conclusions can be applied to spin-1~\citep{East2017vector, Frolov2018vector, Siemonsen2019vector} and spin-2~\citep{deRham2014spin2, Hinterbichler2011spin2} bosons if efficient black hole spindown is possible.}
We summarize our semi-analytic model for studying black hole mergers in dense star clusters \citep[based on][]{Antonini2018hierarchical}, in Sec.~\ref{sec:model}.
We present results from cluster simulations in relation to both individual clusters and the population as a whole in Sec.~\ref{sec:results}.
Finally, we outline the implications of superradiance for recently observed black hole mergers, and the possibility for the detection of ultralight bosons in this boson mass regime in Sec.~\ref{sec:implications}.

\section{Black Hole Superradiance}\label{sec:sr}

In this section we summarize the key expressions from~\cite{Brito2020superradiance} to provide the relevant theory for including the effects of ultralight bosons in cluster simulations. 
We work with the analytic approximations for the evolution of boson clouds around spinning black holes~\citep{Ternov1978scalar, Detweiler1980unstable, Baryakhtar2017rem}, as opposed to the coupled differential equations governing the black hole-boson cloud system assuming quasi-adiabatic evolution~\citep{Brito2015Super}.
Though these approximations are accurate in the limit $\alpha \lesssim 0.1$, the errors are insignificant when extrapolated~\citep{Pani2012}.

\subsection{The superradiant condition}

The superradiant condition is simply that the boson's angular frequency, $\omega = \mu/\hbar$, must satisfy~\citep{Bekenstein1973SuperA, Brito2020superradiance}
\begin{equation}
    \frac{\omega}{m} < \Omega_\textrm{BH}\equiv \frac{1}{2}\frac{c^3}{GM}\frac{\chi}{1+\sqrt{1-\chi^2}},\label{eq:sr1}
\end{equation}
where $m$ is the magnetic quantum number corresponding to a specific boson cloud mode and $\Omega_\textrm{BH}$ is the angular frequency of the black hole's outer horizon~\citep{Teukolsky2014Kerr}. 
Furthermore, the black hole's angular frequency is a function of the its mass, $M$, dimensionless spin, $\chi$, and dimensionless outer radius,  $\bar{r}_+\equiv r_+/r_\textrm{BH} =1+\sqrt{1-\chi^2}$, where $r_\textrm{BH} = GM/c^2$ where the characteristic black hole length scale which is half the Schwarzschild radius. 
The energy eigenstates of the boson cloud take a similar form to those of a hydrogen atom, and are denoted by radial $n$, orbital $l$, and magnetic quantum numbers~\citep{Ternov1978scalar, Detweiler1980unstable}.

To highlight this comparison, we define the \textit{effective fine-structure constant} for the ``black hole atom'' as the ratio of the characteristic length scale to the boson's Compton wavelength, $\lambdabar = \hbar c/\mu$,
\begin{equation}
    \alpha = \frac{r_\textrm{BH}}{\lambdabar} \equiv \frac{GM\mu}{\hbar c^3}.
\end{equation}
Large values of $\alpha$ lead to significant growth of the boson cloud. 
If the inequality in Eq.~\eqref{eq:sr1} is satisfied, the boson cloud extracts rotational energy from the black hole, leading it to spin down. 
Unstable growth of the boson cloud ceases when the angular frequency of the bosonic wave equals $m\Omega_\textrm{BH}$. 

In order to incorporate the effect of superradiance, we compute the final mass and spin of a black hole according to Eq.~\eqref{eq:sr1}~\citep{Isi2018direct, Brito2017search},
\begin{align}
    M_f &\approx M_i\Big(1-\frac{\alpha_i\chi_i}{m}\Big),\label{eq:Mf}\\
    \chi_f &= \frac{4\alpha_f m}{4\alpha_f^2+m^2}.\label{eq:chif}
\end{align}
Here the subscripts $i$ and $f$ refer to the initial and final states of the boson cloud-black hole system, respectively, if the boson cloud is given enough time to saturate a given mode. 
However, the cloud saturates fully only if $\lambdabar \sim r_\textrm{BH}$. Otherwise the cloud does not grow appreciably during the lifetime of the black hole.

\subsection{Superradiance timescales and evolution}
The occupation number of a given energy state grows exponentially at the rate~\citep{Ternov1978scalar, Detweiler1980unstable, Baryakhtar2017rem},
\begin{equation}
    \Gamma_{jlmn} \approx 2\alpha^{2j+2l+5}\bar{r}_+(m\Omega_\textrm{BH}-\omega)C_{jlmn}, \label{eq:gamma}
\end{equation}
where $C_{jlmn}$ is a dimensionless factor dependent on the quantum state inhabited by the ultralight bosons, and $j$ is the total angular momentum quantum number. In the case of scalar ultralight bosons, $j=l$, and~\citep{Ternov1978scalar, Detweiler1980unstable} 
\begin{align}
    C^\textrm{(scalar)}_{jlmn} &= \frac{2^{4l+2}(2l+n+1)!}{(l+n+1)^{2l+4}n!}\Big(\frac{l!}{(2l)!(2l+1)!}\Big)^2\nonumber\\ 
    &\times\prod_{k=1}^l\Big(k^2(1-\chi^2) + \frac{4r_+^2}{c^2}(m\Omega_\textrm{BH}-\omega)^2\Big).
\end{align}
The growth rate of the occupation number can be used to approximate the timescale for the size of the cloud to grow by one $e$-fold, often known as the \textit{instability timescale}. 

Typically, the size of the boson cloud saturates at $\sim180$ $e$-folds of the instability timescale~\citep[cf. Eq. 14 from][]{Baryakhtar2017rem}, which we use to compute an approximate \textit{growth timescale} around a black hole~\citep{Arvanitaki2010Stringb, Arvanitaki2010String, Baryakhtar2017rem, Ng2019constraints},
\begin{equation}
    \tau^{(\textrm{scalar})}_{jlmn} \approx \frac{180}{\Gamma_{jlmn}}.\label{eq:growthtimescale}
\end{equation}
The growth timescale scales as
\begin{equation}
    \tau^{(\textrm{scalar})}_{jlmn} \propto M \alpha^{-4\ell -5},
\end{equation}
where it is clear that when $\alpha \ll 0.1$, the instability growth rate is greatly reduced~\citep{Brito2015Super}. 

Finally, due to the greatly differing timescales between states, the mode which determines the final spin of the black hole is the mode with the lowest final spin that grows within the age of the black hole, or \textit{evolution timescale}, $\tau_\textrm{evol}$ (e.g. see Fig. 1 in \cite{Ng2019constraints}, Fig. 3 in \cite{Baryakhtar2017rem}).
Motivated by the strong fine-structure constant dependence, and the associated strong mass dependence, the quasi-adiabatic differential equations governing the evolution of the black hole-boson cloud system can be largely ignored. 
We consider a black hole to have undergone superradiance only if its age exceeds the growth timescale of the fastest growing mode. 
For the remainder of this manuscript, we focus on the first three $l=m=1,2,3$, $n=0$, states. 

\begin{figure*}[t!]
    \centering
    \includegraphics[width=\linewidth]{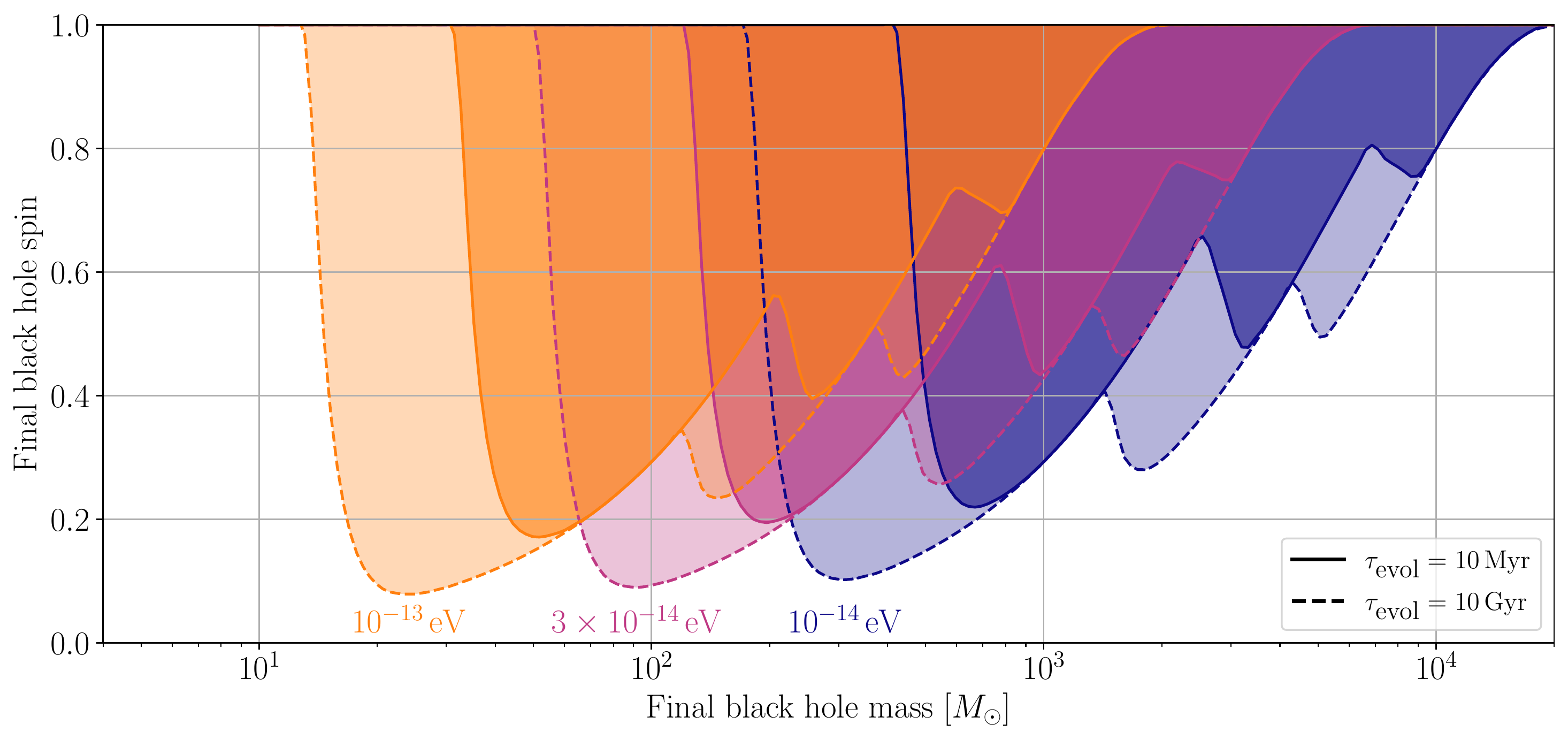}
    \caption{Exclusion regions for black holes in the presence of three different scalar ultralight bosons, with masses $10^{-13}$, $3\times10^{-13}$, and $10^{-14}$\,eV~\citep[cf. Fig 1 from][]{Brito2017search}. The darker-coloured regions bounded by the solid curves correspond to black hole mass-spin parameter space within which any black hole would be spun down via superradiance after $\tau_\textrm{evol} = 10$\,Myr. 
    Since binary black hole merger delay time distribution has little support for systems merging within $10$\,Myr of formation~\citep[e.g.][]{Britt2021dtg}, we expect all black holes to allow for superradiant cloud growth for at least 10\,Myr. 
    Therefore, the observation of a black hole within the darker region of the parameter space can rule out some range of boson mass.
    The lighter region bounded by the dashed curves corresponds to when the black holes are given $10$\,Gyr to evolve.}
    \label{fig:exclusion}
\end{figure*}

An important corollary of Eqs.~\eqref{eq:chif}--\eqref{eq:growthtimescale} is that there are regions of the mass-spin parameter space, also known as the Regge plane, where a black hole cannot exist if an ultralight boson is present~\citep{Brito2015Super, Brito2017search, Arvanitaki2010Stringb}. These \textit{exclusion regions} are strongly dependent of the boson's mass and the evolution timescale of the black hole population.
In Fig.~\ref{fig:exclusion}, we present the exclusion regions as governed by the first three $l=m=1,2,3$, $n=0$ energy eigenstates over the black hole mass range relevant for hierarchical growth from stellar mass black holes.
Each $m$ corresponds to a particular ``bump'' in the excluded region, with the $m=1$ mode contributing at lower black hole masses. 
Furthermore, increasing the evolution timescale of the systems results in the growth of boson clouds around lower mass black holes. 

\section{Semi-analytic cluster model}~\label{sec:model}

To model the evolution of dense stellar clusters such as nuclear star clusters, we follow \cite{Antonini2018hierarchical} which applies H\'enon's principle~\citep{Henon1975} to simulate the evolution of dense stellar environments. 
In this section, we briefly summarize the key components of the cluster model and the evolution of the cluster's black hole population. Please refer to App.~\ref{app:a} for the full details of the model. 

\subsection{Evolution of cluster properties}

To model the global properties of a cluster, we assume the heating rate from black hole binaries in the core of the cluster balances the energy flow into the whole cluster, known as H\'enon's principle~\citep{Henon1961, Henon1975, Gieles2011, Breen2013, Kremer2019}. The half-mass radius, heating rate, and escape velocity of the cluster evolve as
\begin{align}
    \rh(t) &= r_{\textrm{h},0} \Big(\frac{3}{2}\frac{\zeta(t-t_0)}{\tau_{\textrm{rh},0}} + 1\Big)^{2/3},\label{eq:rht}\\
    \dot{E}(t) &= \dot{E}_0 \Big(\frac{3}{2}\frac{\zeta(t-t_0)}{\tau_{\textrm{rh},0}} + 1\Big)^{-5/3},\label{eq:edott}\\
    v_{\textrm{esc}}(t) &= v_{\textrm{esc},0} \Big(\frac{3}{2}\frac{\zeta(t-t_0)}{\tau_{\textrm{rh},0}} + 1\Big)^{-1/3}.\label{eq:vesct}
\end{align}
Here $t_0$ is the time at which the first black holes begin to heat the cluster, and $\zeta\simeq0.1$~\citep{Gieles2011, Alexander2012} is a dimensionless factor.
The initial half-mass radius, heating rate and escape velocity are dependent only on the cluster mass, $M_\textrm{cl}$ and initial density, $\rho_{\textrm{h},0}$, 
\begin{subequations}~\label{eq:initial}
\begin{align}
    &\dot{E}_0 \simeq 2.3\times10^5\,M_\odot\,(\textrm{km s}^{-1})^2\,\textrm{Myr}^{-1}M_5^{2/3}\rho_{5,0}^{5/6},\\
    &\tau_{\textrm{rh},0} \simeq 7.5\,\textrm{Myr}\,M_5\rho_{5,0}^{-1/2},\\
   &v_{\textrm{esc},0} \simeq 50\,\textrm{km s}^{-1}\,M_5^{1/3}\rho_{5,0}^{1/6},
\end{align}
\end{subequations}
where $M_5 \equiv \mcl/10^5\,M_\odot$ and $\rho_{5,0} \equiv \rho_{\textrm{h},0}/10^5\,M_\odot\,\textrm{pc}^{-3}$, and $\ph = 3\mcl/8\pi \rh^3$. 
All the quantities are dependent on only the initial density of the cluster and the cluster mass. 

\subsection{Evolving the black hole population}

Within our cluster model, we must first create a first generation black hole population. 
First generation black holes' masses are assumed to follow the inferred {\tt \sc Power Law + Peak} model~\citep{mass} from the second gravitational-wave transient catalog from the LIGO/Virgo Collaboration~\citep{GWTC2pop}, with an additional strict mass cut-off at $45\,M_\odot$. This model is a combination of a power-law describing low mass black holes, and a Gaussian peak at higher masses motivated by the prediction of a pair-instability supernova upper mass-gap precluding the formation of black holes with masses between $\sim45\,M_\odot$ and $\sim130\,M_\odot$~\citep{Barkat1967, Fryer2000, Heger2002, Belczynski2016, Spera2017, Stevenson2019}.

All 1G black holes are considered to be initially non-spinning, motivated by studies which indicate that isolated black holes are likely to form with small spins ($\chi \lesssim 0.01$)~\citep{Fuller2019}.
The total mass of the first generation black holes within the cluster is fixed to $2\%$ of the cluster mass~\citep{Lockmann2010, Antonini2018hierarchical, Kremer2020}.
Additionally, each black hole is initialized with a natal supernova kick velocity~\citep{Hobbs2005, Fryer2001}, though this has a minimal effect on the initial population. 
Finally, the black holes are deposited within the cluster at the initial half mass radius, $r_{\textrm{h},0}$. 
Each black hole is expected to settle within the core of the cluster on the dynamical friction timescale, $\tau_{\textrm{df}}(r_{\textrm{h},0}, t_0)$~\citep{binney2011galactic}, 
\begin{equation}
    \tau_{\textrm{df}}(r, t) \simeq 0.346\frac{r^2 v_\textrm{esc}(t)}{G M\ln\Lambda},~\label{eq:df}
\end{equation}
where $\ln\Lambda\simeq 10$ and assumes the \cite{King1966} cluster model to relate the cluster's velocity dispersion to escape velocity. 

Once the first black holes settle within the cluster's core, a black hole binary might form through dynamical three-body interactions~\citep{Ivanova2005,Morscher2015,Park2017}. 
We let only one black hole binary exist at any one time such that it dominates the heating of cluster.
The binary is formed according to a mass distribution given by $\propto (M_1+M_2)^4$~\citep{OLeary2016}.  
The required heating rate of the cluster is balanced with the loss of energy from the binary in the core of the cluster~\citep{Antonini2018hierarchical, Kremer2019}.
The timescale during which dynamical interactions dominate the energy flow of the cluster is given by 
\begin{equation}
    \tau_\textrm{dyn} \simeq \frac{GM_1M_2}{2a_m}\dot{E}^{-1}(t), ~\label{eq:tdyn}
\end{equation}
where $a_m = \max(a_{\textrm{ej}}, a_{\textrm{GW}})$. Here $a_\textrm{ej}$ is the binary separation at which the binary is ejected, and $a_\textrm{GW}$ is the separation at which gravitational-wave radiation dominates. Since we are focusing on denser stellar environments, the majority of binary systems are likely to merge within the cluster rather can be ejected. The ejection of the interloper black holes is also incorporated~(see App. \ref{app:interloper}).
We calculate the separation at which gravitational-wave radiation dominates by equating the separation evolution due to dynamical interactions and gravitational-wave emission~\citep{Peters1964}.
We compute the timescale for the binary to merge due to gravitational-wave radiation as $\tau_\textrm{GW} = a_m/|\dot{a}_\textrm{GW}|$. 
At this point in the binary's evolution, we modify the black holes' masses and spins according to Eqs.~\eqref{eq:chif} and~\eqref{eq:Mf} to incorporate the effects of superradiance, ensuring each black hole's age exceeds to the growth timescale of the boson cloud. 

\subsection{Black hole merger remnants}

A vital component to the semi-analytic model is the computation of the black hole merger remnant properties. 
In particular, due to the conservation of linear momentum, the final remnant black hole receives a kick from the anisotropic emission of gravitational waves~\citep{Campanelli2006kick, Campanelli2007kick, Campanelli2007kickB, Gonzalez2007kick, Gonzalez2007kickB, Lousto2010kick, Lousto2012kick, Varma2018kick, Merritt2004conseq, Varma2020extractingKick}.
This recoil velocity, $v_k$, determines whether the remnant remains in the core, is ejected from the core and has to settle through dynamical friction, or is ejected from the cluster entirely.
We utilize the {\tt \sc Precession} code~\citep{Gerosa2016} to determine the final mass~\citep{Barausse2012}, spin~\citep{Barausse2009}, and recoil velocity of the remnant black hole. 
The details of the calculation of the recoil velocity are outlined in App.~\ref{app:kick}.

The salient features of the recoil velocity calculation for the study of hierarchical mergers arise from black hole binary spins and mass ratios. 
It has been found that the largest kicks are typically a result of special spin configurations known as \textit{superkicks}~\citep{Campanelli2007kick, Gonzalez2007kickB} and \textit{hang-up kicks}~\citep{Lousto2011, Lousto2012kick}. 
These spin configurations can lead to recoil velocities up to $\sim 5000$\,km\,s$^{-1}$.
Conversely, non-spinning binaries typically lead to the smallest recoil velocities. 
For example, the largest recoil velocity a non-spinning binary can produce is only $\sim 170$\,km\,s$^{-1}$ for $q\simeq1/3$~\citep{Gonzalez2007kick}. 
In order to quantify the binary's total spin and explore its contribution to hierarchical mergers, we introduce a mass-weighted spin magnitude,
\begin{equation}
    \langle\chi\rangle \equiv \frac{\chi_1 + q^2\chi_2}{(1+q)^2} = \frac{M_1^2\chi_1+M_2^2\chi_2}{(M_1+M_2)^2}.\label{eq:mspin}
\end{equation}

After calculating the properties of the merger remnant, we determine if or where it should be re-introduced into the black hole population. 
If the recoil velocity exceeds the escape velocity of the cluster at the time of the merger, $v_k \geq v_\textrm{esc}(t_\textrm{merge})$, then the remnant is ejected from the cluster. 
Alternatively, if $v_k < v_\textrm{esc}(t_\textrm{merge})$, the remnant is retained and placed at a radius~\citep{Antonini2018hierarchical}
\begin{equation}
    r_{\textrm{in}} \simeq r_\textrm{h}(t_\textrm{merge}) \sqrt{\frac{v_\textrm{esc}^4(t_\textrm{merge})}{(v_\textrm{esc}^2(t_\textrm{merge}) - v_k^2)^2} - 1}.
\end{equation}
If $r_\textrm{in} < r_\textrm{h}(t_\textrm{merger})/10$, a conservative estimate of the core radius~\citep{Hurley2007, Madrid2012}, then the remnant black hole remains in the core.
Otherwise, the binary must first settle in the core due to dynamical friction (cf. Eq.~\eqref{eq:df}). 
After the merger remnant has either been ejected (over a period $\tau_\textrm{dyn}$) or retained (over a period of $\tau_\textrm{dyn} +\tau_\textrm{GW}$) either in the core or in the cluster, we repeat process outlined above by generating a new black hole binary to support the heating rate condition. 
The simulation is concluded when either only one black hole remains, too few black holes remain for dynamical hardening, or 13\,Gyr have passed. 

\section{Results}~\label{sec:results}

We create a grid of $\approx 1800$ simulations in the space of cluster mass $\mcl$, initial density $\rho_{\text{h},0}$, and boson mass $\mu$.
We select log-uniformly spaced cluster masses between $10^6$ and $2\times10^8\,M_\odot$, initial densities between $10^3$ and $10^7\, M_\odot$\,pc$^-3$, and  boson masses between $5\times10^{-15}$ and $5\times10^{-13}$\,eV. 
We run the simulations with and without ultralight bosons.\footnote{For each point in the grid, we run 30 sub-simulations to average over random fluctuations.}

\subsection{Individual cluster simulations}\label{subs:indiv}

\begin{figure*}[p!]
    \centering
    \includegraphics[width=\linewidth]{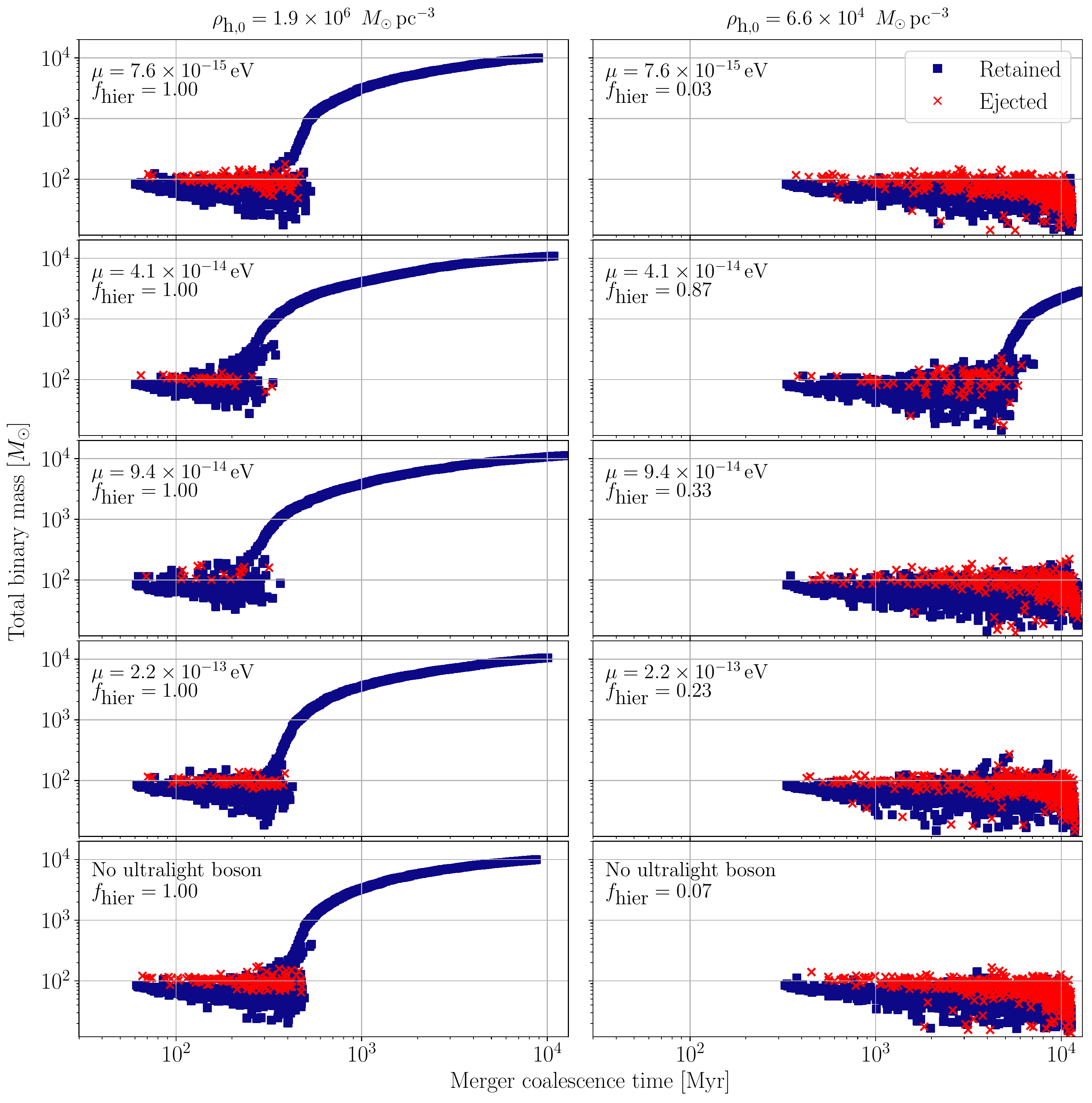}
    \caption{Total binary black hole merger mass versus the coalescence time of the merger from simulations with $M_\textrm{cl} = 1.1\times10^7\,M_\odot$ and two different initial densities. 
    From top to bottom, each row is an increasingly large boson mass except the bottom row, which shows the results when no boson is present.
    The two columns show how the results change for different initial densities.
    The shape of the points corresponds to whether the merger remnant was ejected (red cross) or retained (blue square). 
    Two distinct phases of dynamical interactions and hierarchical black hole growth are present.
    The denser cluster (left) allows for all clusters to dynamically form heavy black holes ($M > 10^3\,M_\odot$).
    The inclusion of ultralight bosons with $\mu = 4.1\times10^{-14}$\,eV leads to hierarchical black hole growth in less dense cluster where growth does not occur under normal circumstances (right).
    The fractions of simulations leading to hierarchical black hole growth ($f_\textrm{hier}$) are also stated. 
    }
    \label{fig:timeevolution}
\end{figure*}

In Fig.~\ref{fig:timeevolution} we plot the total mass of the binary black hole system as a function of the coalescence time. 
These merger evolution tracks are plotted for four different ultralight boson masses (and for the case of no ultralight boson) using two different initial cluster densities. 
The shape of the points contained to retained (blue square) or ejected (red cross) binary systems.
Different initial densities lead to different escape velocities and dynamical friction timescales. 
The denser cluster (left), with an initial density of $\rho_{\textrm{h},0} = 1.9\times10^6\,M_\odot$\,pc$^{-3}$, has an initial escape velocity of $v_{\textrm{esc},0} = 391.8$\,km\,s$^{-1}$. 
Whereas, the less dense cluster (right) only has an initial escape velocity of $v_{\textrm{esc},0} = 224.2$\,km\,s$^{-1}$. 

There are two distinct epochs during a black hole population's evolution in a cluster. 
Initially, the black hole population undergoes random, dynamical interactions which lead to mergers and formation of 2G or third-generation (3G) black holes.
Since the black hole binaries formed early within the cluster have similar total masses, the chance any two black holes become a binary is small, though heavier black holes are slightly more likely to form binaries~\citep{OLeary2016}. 
This epoch is categorized by remnant masses typically less than $\sim200\,M_\odot$ in the early cluster evolution. 
Eventually, however, one black hole tends to become sufficiently massive to dominate, leading to the second evolutionary phase.
Since the heavy black hole will primarily form binaries with much smaller black holes, the resulting small mass ratio can lead to significantly reduced gravitational-wave recoil velocities~\citep{Gonzalez2007kick}.
This black hole therefore seeds hierarchical growth in the cluster.
This period of evolution is seen by a track of increasing total binary mass with the merger coalescence time in Fig.~\ref{fig:timeevolution}. 

In the left column of Fig.~\ref{fig:timeevolution}, we show the binary black hole mergers which occurr in a cluster simulation with $M_\textrm{cl}=1.1\times10^7\,M_\odot$ and $\rho_{\textrm{h},0} = 1.9\times10^6\,M_\odot$pc$^{-3}$. 
We see that all simulations are able to support hierarchical black hole growth. However, the existence of ultralight boson at some masses can change the time at which hierarchical growth starts to occur. 
This is due to the spin down of 2G+1G and 2G+2G generation black hole mergers (cf. Fig.~\ref{fig:exclusion}) when bosons of masses $\sim2\times10^{-14}$ -- $2\times10^{-13}$\,eV exist helping retain more merger remnants.
By retaining more massive black holes, the formation of a hierarchical black hole seed is more likely to occur earlier in the simulation.
The simulation result in the top panel ($\mu = 7.6\times10^{-15}$\,eV) is similar to the case when no ultralight boson exists because the boson cloud instability growth rate is significantly reduced for stellar mass black holes. 
However, there is some indication of spin-down from superradiance in the high mass, hierarchical growth regime. 

On the right of Fig.~\ref{fig:timeevolution}, we show the results from a cluster with $M_\textrm{cl}=1.1\times10^7\,M_\odot$ and $\rho_{\textrm{h},0} = 6.6\times10^4\,M_\odot$\,pc$^{-3}$.  
Only one particular ultralight boson, with a mass of  $\mu = 4.1\times10^{-14}$\,eV, can facilitate hierarchical black hole growth.
The clear difference between this simulation and the remaining four is that the binary systems with a total mass $\sim 80$ -- $200\,M_\odot$ have mass-weighted spins $\langle \chi \rangle \lesssim 0.15$.
\textit{\textbf{The important result from these individual simulations is that the presence of ultralight bosons can impact what cluster properties support hierarchical growth}}. 
For the remainder of the manuscript, we use the scenario where $\mu = 4.1\times10^{-14}$\,eV as our best-case scenario example.

\begin{figure*}[t!]
    \centering
    \includegraphics[width=0.48\linewidth]{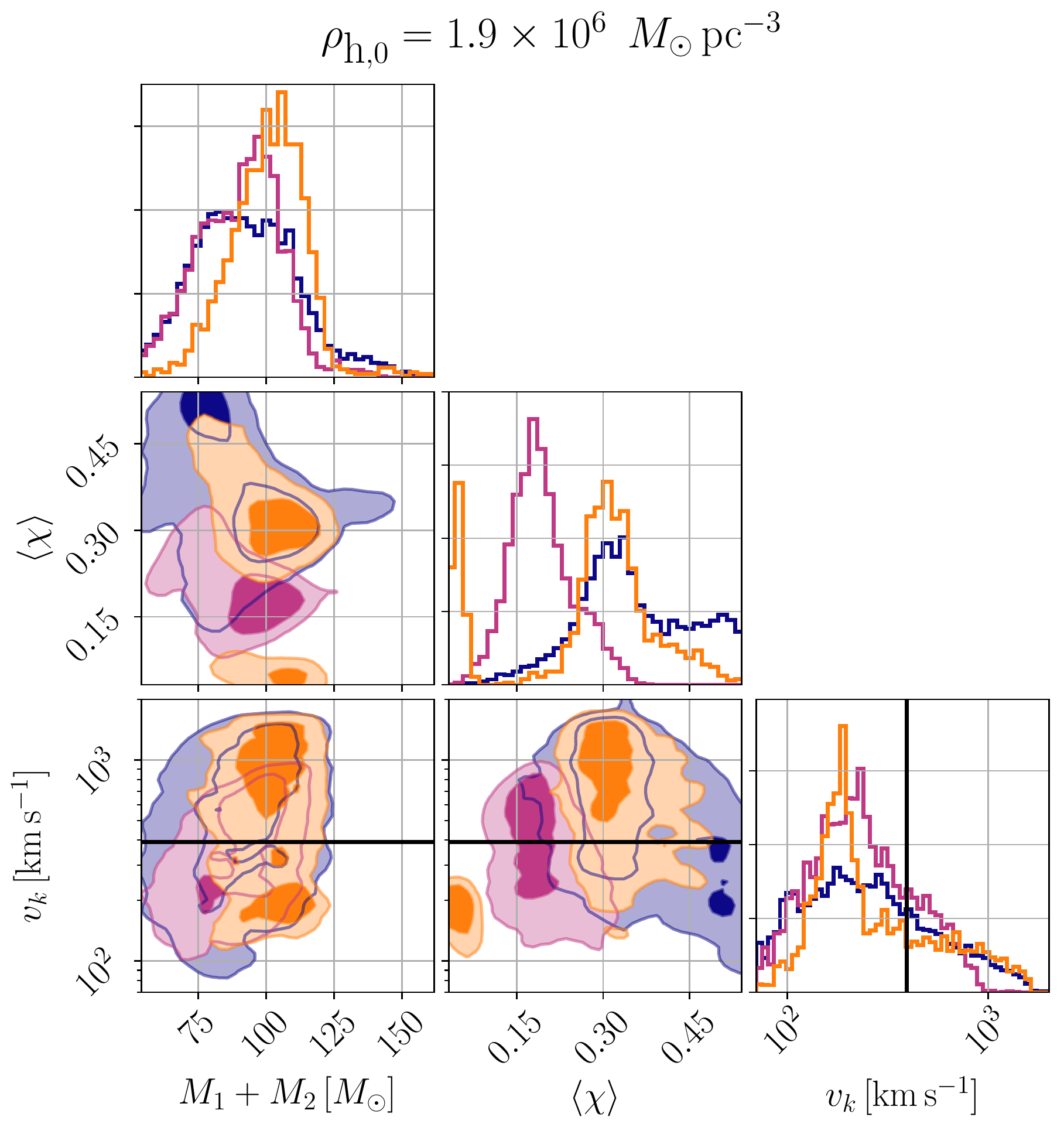}
    \includegraphics[width=0.48\linewidth]{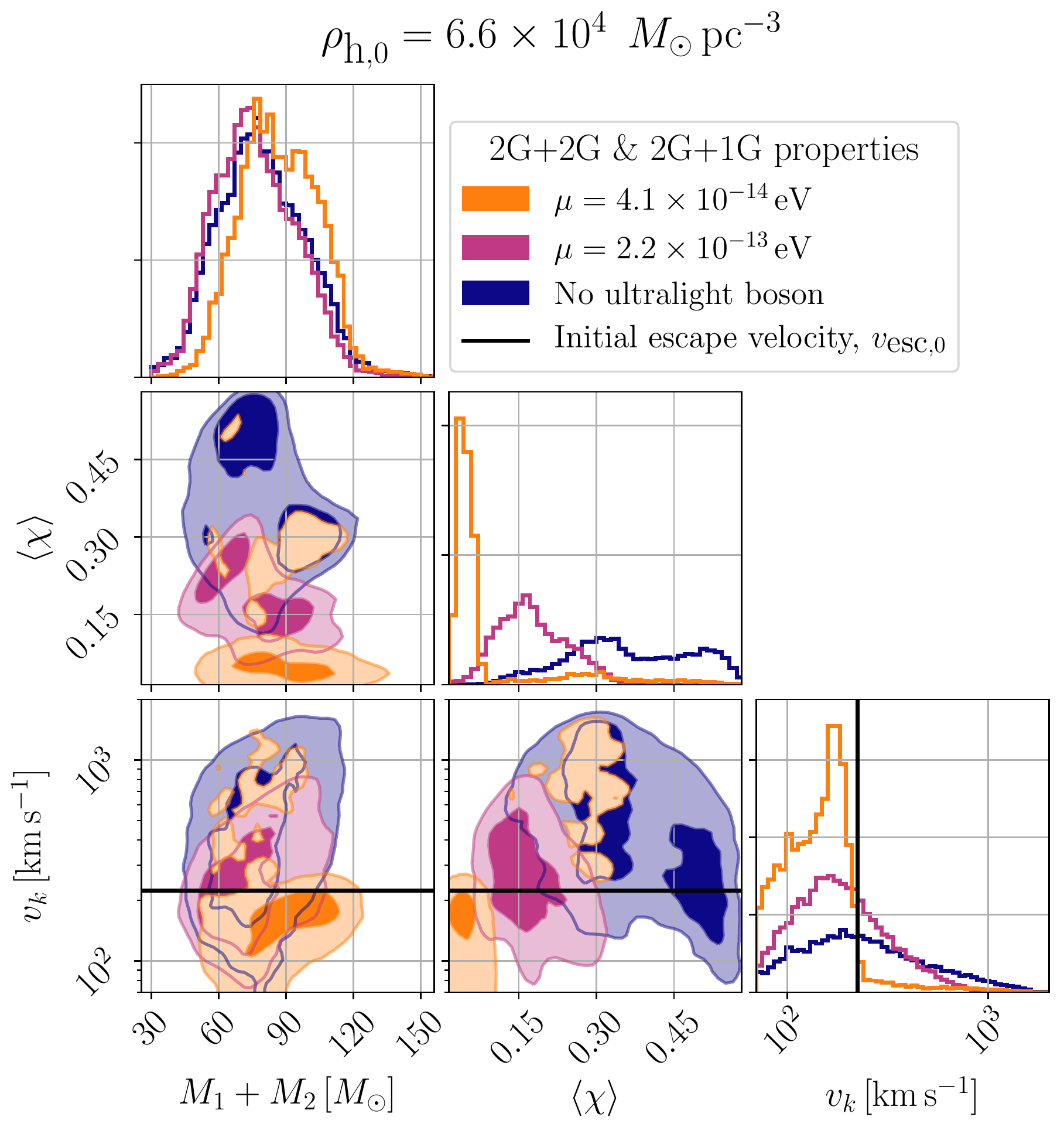}
    \caption{
    Distributions of the total binary black hole mass, mass-weighted spin ($\langle \chi \rangle$), and gravitational recoil velocity ($v_k$) for all binaries forming a third generation (3G) black hole, two second generation black holes (2G+2G) and second generation and first generation (2G+1G) merger events.
    The black lines corresponds to the initial escape velocity of the cluster, $v_{\textrm{esc,}0}$.
    The distributions correspond to six panels in Fig.~\ref{fig:timeevolution}, when $\mu = 4.1\times10^{-14}$, $2.2\times10^{-13}$\,eV and no ultralight boson is present. Of these ultralight boson masses, only $\mu = 4.1\times10^{-14}$\,eV (orange) facilitates hierarchical black hole growth in both example clusters considered. 
    Therefore, black holes spinning down due to superradiance directly leads to a higher retention fraction of black holes and consequently a higher chance of hierarchical growth.
    }
    \label{fig:corner}
\end{figure*}

In Fig.~\ref{fig:corner}, we plot the distributions of the binary black hole merger properties for all 2G+2G and 2G+1G mergers for cluster simulations presented in Fig.~\ref{fig:timeevolution}.
We use the full set of 30 simulations for each set of initial cluster parameters, and show the distributions for the $\mu = 4.1\times10^{-14}$\,eV, $2.2\times10^{-13}$\,eV and no ultralight boson simulations.
The left plot shows the distributions in the case where all clusters lead to hierarchical growth. 
The bulk of the kick velocity distributions lay below the initial escape velocity, $v_{\textrm{esc},0}$, of the cluster. Crucially, in the $\mu=4.1\times10^{-14}$\,eV results (orange), we see two spin populations corresponding to systems which have undergone substantial superradiance ($\langle\chi \rangle \lesssim 0.15$), and a population which has not ($\langle\chi \rangle \sim 0.3$--$0.45$). The low-spin population is entirely retained within the cluster.

In the right panel, only the $\mu=4.1\times10^{-14}$\,eV (orange) distribution corresponds to a system capable of hierarchical growth. 
In these results, the majority of black holes are spun down to the low-spin population, and are still retained. 
Conversely, in the cases of $\mu=2.2\times10^{-13}$\,eV and no ultralight boson, the black holes are not spun-down enough to reduce the velocity distribution significantly.
Therefore the majority of second-generation black holes are ejected upon merging with another black hole. 

\subsection{Hierarchical growth in present-day clusters}\label{subs:present}

\begin{figure}[t!]
    \centering
    \includegraphics[width=\linewidth]{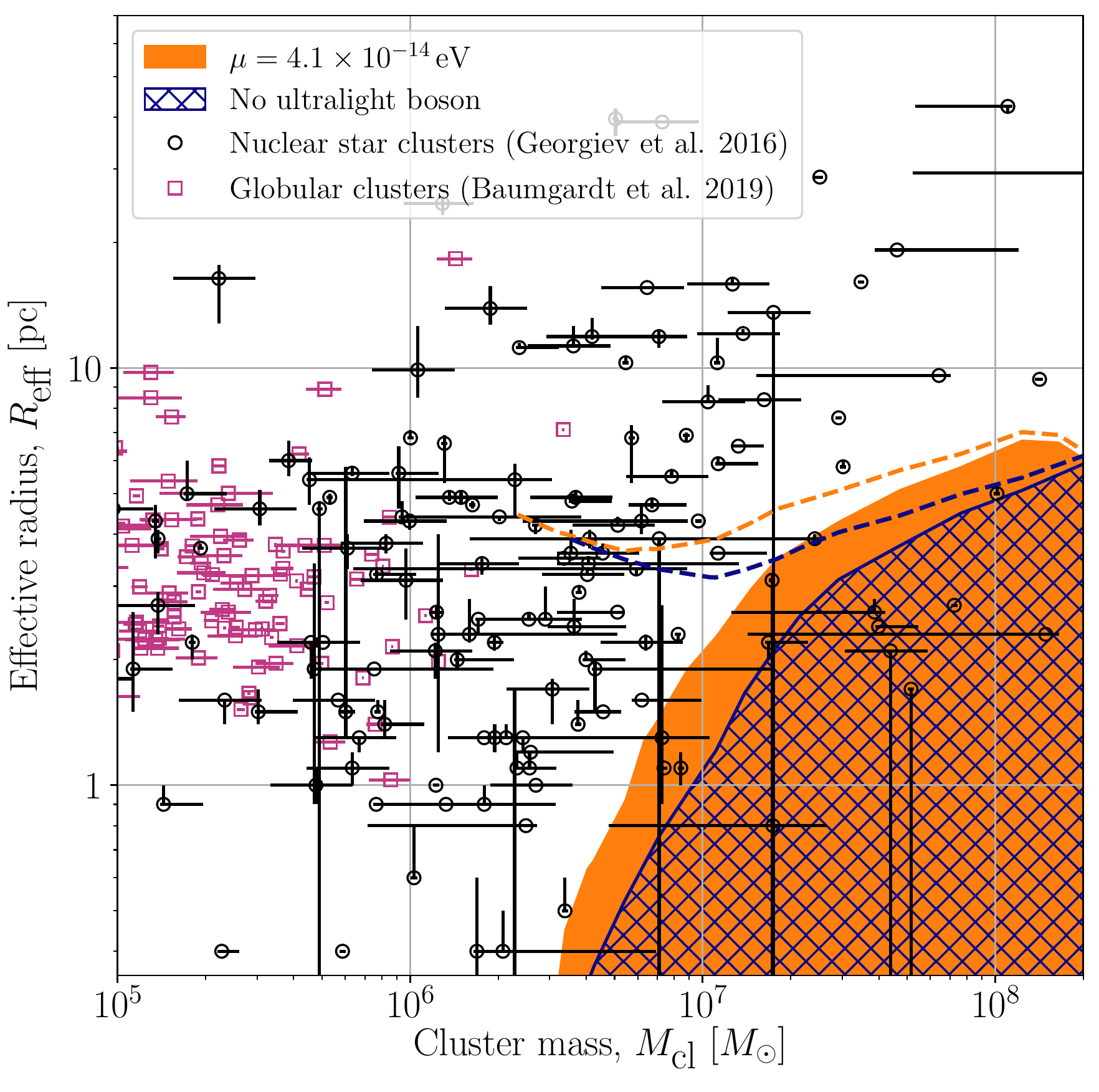}
    \caption{Effective radius ($R_\textrm{eff}$) --- cluster mass ($M_\textrm{cl}$) parameter space that can support hierarchical growth, and present-day populations for globular clusters~\citep{Baumgardt2019} and nuclear star clusters~\citep{Georgiev2016}. The solid orange and hatched blue contours correspond to the regions of the parameter-space where more than $50\%$ of cluster simulations at a given point can support hierarchical growth if $\mu=4.1\times10^{-14}$\,eV or if no ultralight boson exists, respectively. 
    We assume the effective radius is approximately given by the half-mass radius as $R_\textrm{eff} \simeq 0.75r_\textrm{h}$. 
    Any observed cluster within this contour is capable of undergoing hierarchical black hole growth now or in the future. 
    The dashed contours indicate the evolved state of the bounded region after $10$\,Gyr. 
    A cluster under the dashed contours might have supported hierarchical growth in the two scenarios.
    }
    \label{fig:cluster_properties}
\end{figure}

For each set of simulations with a given ultralight boson mass and cluster properties, we compute the fraction of the repeated 30 simulations which result in the formation of a black hole with a mass $\geq1000\,M_\odot$\footnote{
This threshold is somewhat arbitrary, as we still compute similar fractions when considering a mass threshold as low as $\sim400\,M_\odot$. This is because once a $\sim400\,M\odot$ black hole is formed, hierarchical growth almost always follows.}.
We denote this fraction $f_\textrm{hier}$~\citep{Antonini2018hierarchical}.
We use $f_\textrm{hier}$ as a proxy for the fraction of simulations leading to hierarchical growth within the cluster.
We compute the region within which more than $50\%$ of the simulations for a given initial effective radius $R_\textrm{eff}$ and cluster mass undergo hierarchical growth, i.e., $f_\textrm{hier} > 0.5$. 
We compare the region of the effective radius-cluster mass ($R_\textrm{eff} - M_\textrm{cl}$) parameter space where clusters support hierarchical growth with a population of globular clusters~\citep{Baumgardt2019} and nuclear star clusters~\citep{Georgiev2016}. 
For the nuclear star cluster population, we plot the 152 nuclear star clusters with cluster mass uncertainty less than an order of magnitude, and effective radius estimates with only positive radius support. However, we use the full population of 228 nuclear star clusters from~\citep{Georgiev2016} in future calculations. 
These results are presented in Fig.~\ref{fig:cluster_properties} for $\mu=4.1\times10^{-14}$\,eV (orange), as well as simulations where no ultralight boson is present (hatched blue). 

The lower right region, corresponding to denser and heavier clusters, unsurprisingly supports hierarchical growth. 
Our bounds are similar to those presented in \cite{Antonini2018hierarchical} for the scenario with no ultralight boson, though here we empirically compute the region of parameter space supporting hierarchical growth.
The inclusion of superradiance extends the region within which hierarchical growth is supported further into the astrophysical population of clusters, though it still does not permit hierarchical growth in globular clusters. 
Importantly, the orange contour includes a higher percentage of the nuclear star cluster population. 
Furthermore, the highlighted contour region only corresponds to the initial conditions, whereas it is inevitable that the clusters have evolved since their formation. 
To illustrate this, we evolve clusters with initial conditions along the contour to an age of 10\,Gyr, shown by the dashed curves in Fig.~\ref{fig:cluster_properties}. 
A larger proportion of the nuclear star cluster population is contained under this contour. 
However, given that the age of present-day clusters is not well known, it is difficult to interpret whether hierarchical growth might have previously occurred in observed clusters. 

\begin{figure}[t!]
    \centering
    \includegraphics[width=\linewidth]{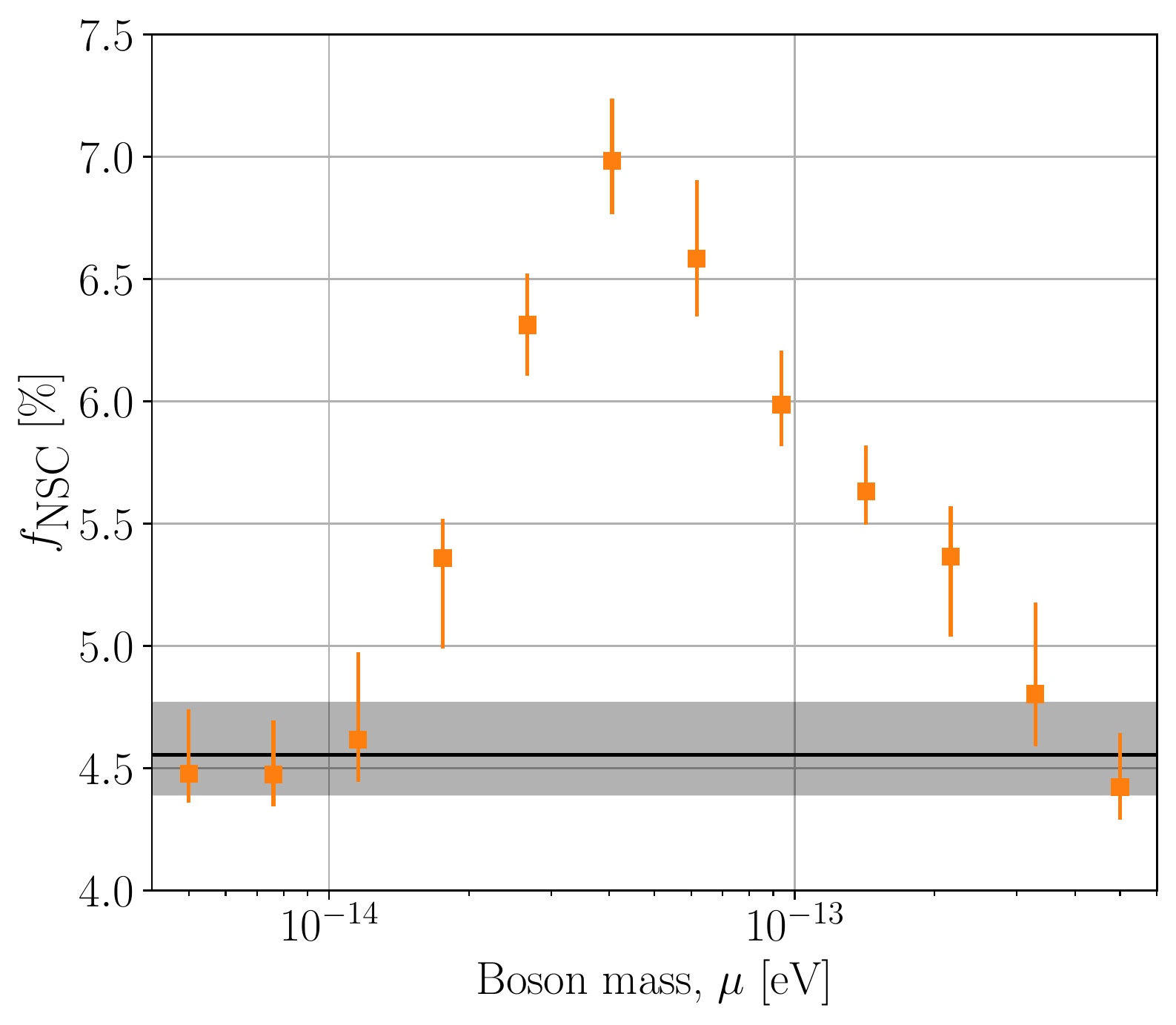}
    \caption{The fraction (and $3\sigma$ uncertainty) of observed nuclear star clusters capable of supporting present-day hierarchical black hole growth under the assumption of the existence of different ultralight bosons. 
    The fraction is computed from the expression in Eq.~\eqref{eq:frac}, using contours such as those presented in Fig.~\ref{fig:cluster_properties}. 
    In the absence of ultralight bosons, we find only $\approx4.5\%$ of observed nuclear star clusters from~\citep{Georgiev2016} can support hierarchical growth (black interval shown), whereas bosons with masses of $\mu \sim 2\times10^{-14}$ --- $2\times10^{-13}$\,eV lead to an increased fraction (shown in orange).}
    \label{fig:fnc}
\end{figure}

In order to interpret the significance of the difference between the hierarchical growth regions for simulations with and without superradiance, we compute the fraction of nuclear star clusters from~\cite{Georgiev2016}, which presently would support hierarchical growth, $f_\textrm{NSC}$, i.e., those nuclear star clusters that lie within the contours in Fig.~\ref{fig:cluster_properties}.
See App.~\ref{app:fnc} for details about the calculation and uncertainty estimation. 
As an example, for the contours presented in Fig.~\ref{fig:cluster_properties}, the fraction of present-day NSCs which can sustain hierarchical growth is $\approx4.5\%$ in the absence of ultralight bosons and $\approx 7\%$ in the presence of a boson with a mass of $\mu = 4.1\times10^{-14}$\,eV. 
The fraction in the absence of ultralight bosons is less than the value presented in \cite{Antonini2018hierarchical} ($f_\textrm{NSC} \simeq 10\%$) as we empirically generate the contour from simulations, as well as explicitly integrate under it to calculate $f_\textrm{NSC}$.
The values of $f_\textrm{NSC}$ for different boson masses are presented in Fig.~\ref{fig:fnc}. The fraction peaks at $\mu\simeq4.1\times10^{-14}\,$eV, corresponding to a $\approx60\%$ increase in the number of clusters capable of supporting hierarchical growth presently. There is also an increased number of clusters currently capable of supporting hierarchical growth if bosons with masses $\mu \sim 2\times10^{-14}$ --- $2\times10^{-13}$\,eV exist. 

\subsection{Synthetic present-day black-hole population}\label{subs:presentpop}

\begin{figure}[t!]
    \centering
    \includegraphics[width=\linewidth]{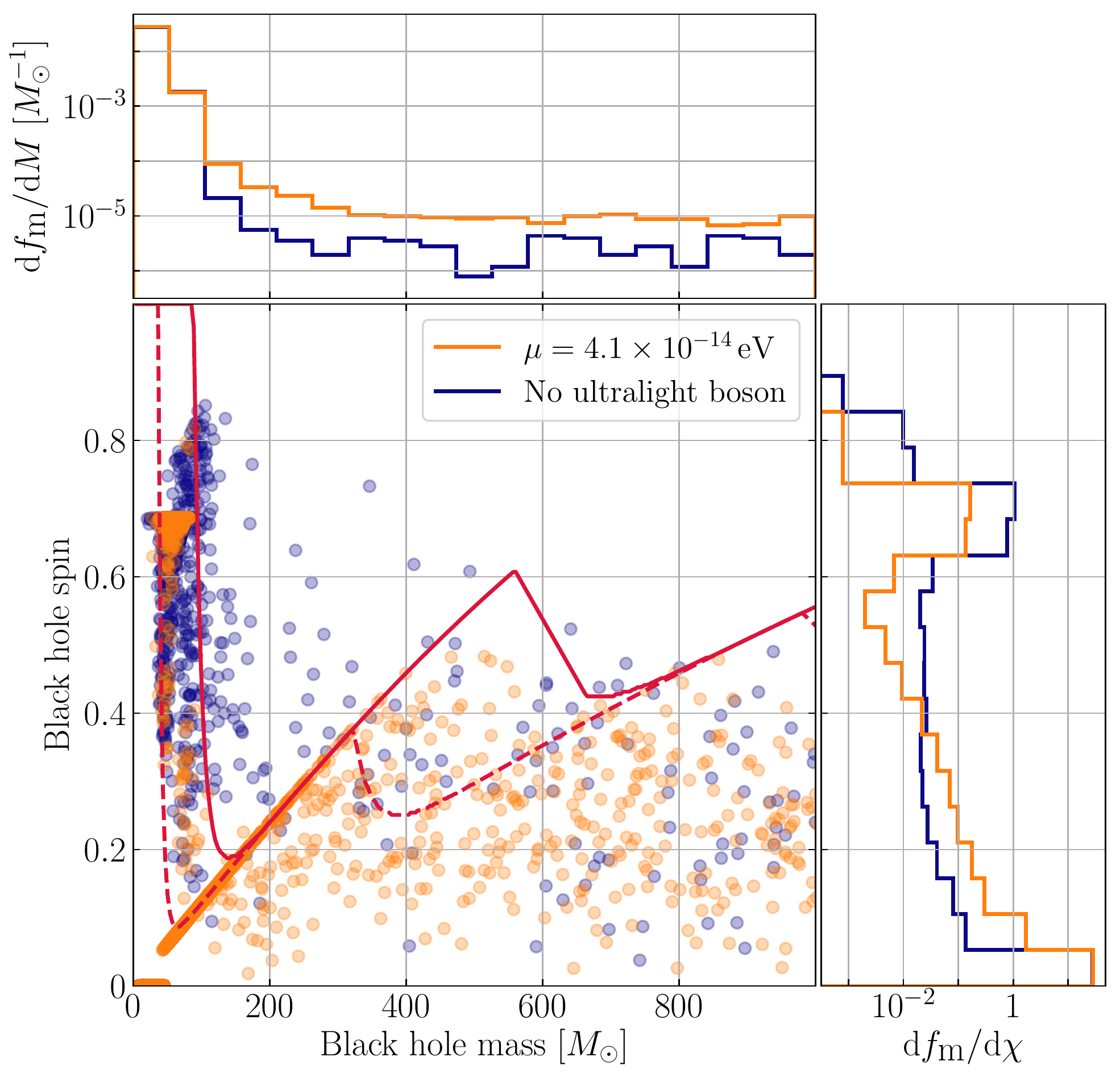}
    \caption{Distribution of the masses and spins of simulated merging black hole populations within a model cluster population visually matching the observed nuclear star cluster population from \cite{Georgiev2016} after evolution, with either an ultralight boson of mass $\mu=4.1\times10^{-14}$\,eV or no ultralight boson.
    The red curves correspond to the $\tau_\textrm{evol} = 10$\,Myr (solid) and $\tau_\textrm{evol} = 10$\,Gyr (dashed) spindown exclusion limits when $\mu=4.1\times10^{-14}$\,eV. 
    With the inclusion of ultralight bosons with $\mu=4.1\times10^{-14}$\,eV, there is an increased number of heavy black holes in the merging population as a direct result of more clusters being able to facilitate hierarchical growth. 
    }
    \label{fig:scatter}
\end{figure}

To understand the distribution of binary black hole mergers in present-day clusters, we generate a synthetic population, which we evolve to a similar state as the observed population. 
We generate cluster masses from $\log_{10}(M_\textrm{cl}/M_\odot) \sim \mathcal{N}(\mu=6.3, \sigma=0.8)$ and  initial densities from $\log_{10}(\rho_{\textrm{h},0}/M_\odot\,\textrm{pc}^{-3})\sim\mathcal{N}(\mu=4.5, \sigma=1.5)$.
Each cluster is evolved to an age drawn at random from $\log_{10}(t_\textrm{age}/\textrm{Gyr}) \sim\mathcal{N}(\mu=0.31, \sigma=1)$. 
These distributions are chosen such that the evolved population visually appears similar to the observed nuclear star cluster properties from \cite{Georgiev2016}. 
We evolve $4.8\times10^4$ clusters with properties drawn randomly from these distributions, for scenarios where $\mu = 4.1\times10^{-14}$\,eV and no ultralight boson exists. 
The final merger is then included in the merger population. 
The individual black hole properties from both scenarios, along with the spin-down limits set by $\mu=4.1\times10^{-14}$\,eV, are shown in Fig.~\ref{fig:scatter}. The densities of the merger fraction, $f_m$, in mass and spin are also presented.

From the one-dimensional marginal distributions, the inclusion of superradiance from bosons with $\mu=4.1\times10^{-14}$\,eV clearly leads to an enhancement of the number of black holes with masses above $100\,M_\odot$ in the population of merging binary systems. 
Furthermore, a distinct 2G population with spins $\sim0.7$ is observed. This is present regardless of the presence of the ultralight boson, as the majority of black holes in this population are not old enough to undergo any significant superradiant boson cloud growth. 
Additionally, the superradiant instability leads to a strong correlation between the black hole spin and mass in the mass range from $\sim80$---$300\,M_\odot$, as the black holes are maximally spun down through the formation of a boson cloud. 
Finally, there is a trend of reduced spins as the black holes become heavier. 
This is due to the fact that these black holes are formed through high mass ratio mergers, which typically reduce the spin of the merger remnant. 
This is present in both scenarios. 

\section{Implications}~\label{sec:implications}

The process of black-hole superradiance can increase the number of nuclear star clusters where hierarchical black hole growth occurs while impacting the black hole merger population.
The signature of this distinct population may be detectable by gravitational-wave detectors such as Advanced LIGO~\citep{aasi2015characterization} and Advanced Virgo~\citep{acernese2014advanced,GW190521a, GW190521b, GWTC2cat}. 

\subsection{Gravitational-wave source population}
\begin{figure*}[t!]
    \centering
    \includegraphics[width=\linewidth]{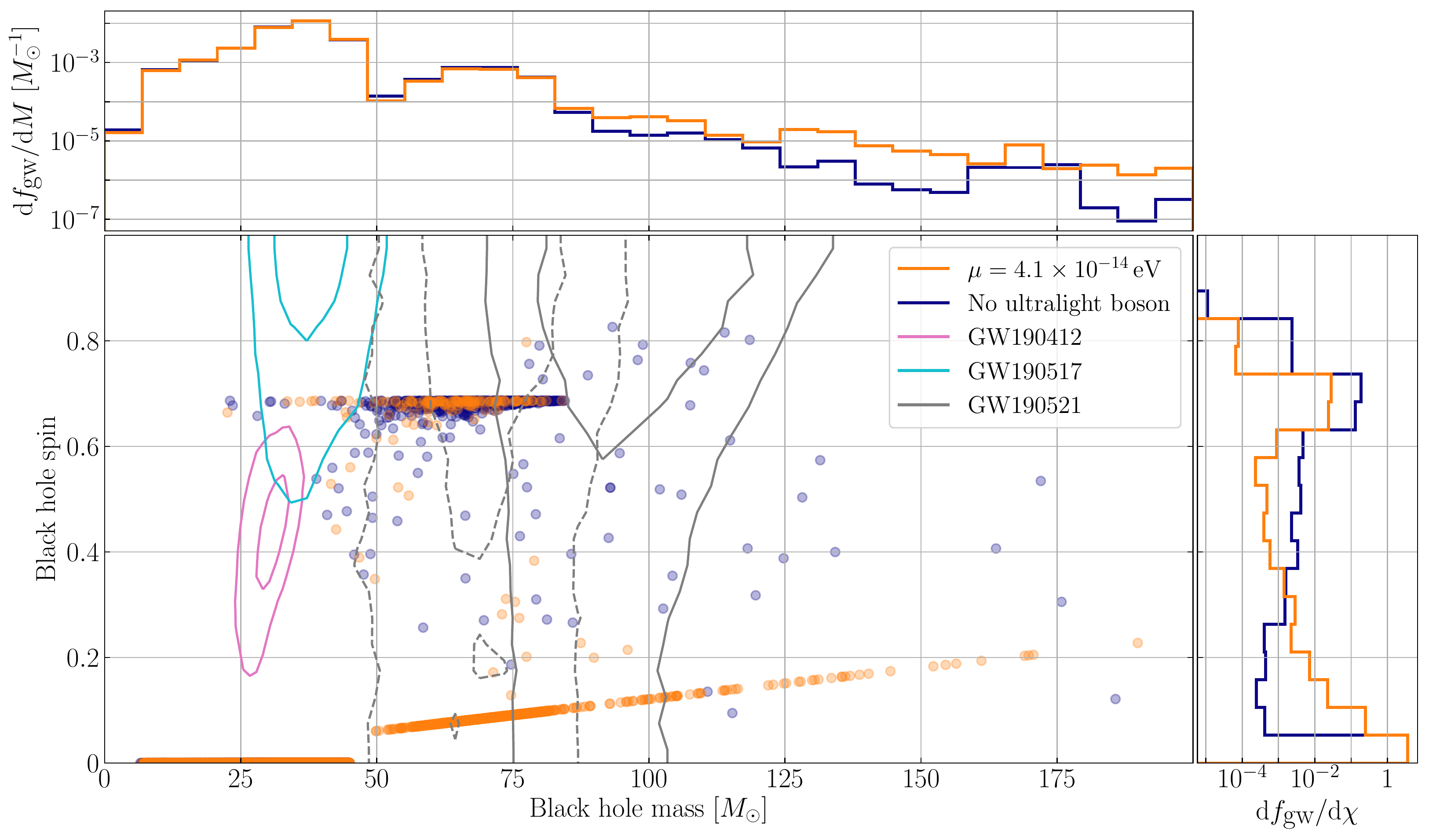}
    \caption{Simulated binary black hole merger population weighted by their detection probability by a single design sensitivity aLIGO detector~\citep{aasi2015characterization} for the $\mu=4.1\times10^{-14}$\,eV or no ultralight boson scenarios. 
    The $1\sigma$ and $2\sigma$ credible intervals inferred for the primary black hole properties from GW190412~\citep{GW190412}, GW190517~\citep{GWTC2cat}, and GW190521~\citep{GW190521a, GW190521b} are shown as solid contours. 
    The properties of GW190521's secondary black hole is given by the dashed contours. 
    The existence of a $\mu = 4.1\times10^{-14}$\,eV ultralight boson would lead to an enhancement of heavier binary systems with lower spins.
    Currently observations from gravitational waves cannot rule out the existence of ultralight bosons in the mass range $10^{-14}$ --- $10^{-13}$\,eV.}
    \label{fig:gwpop}
\end{figure*}

In Fig.~\ref{fig:gwpop}, we plot the mass-spin distribution while taking into account gravitational-wave detector selection biases. 
The primary black hole mass-spin $1\sigma$ and $2\sigma$ two-dimensional credible intervals for GW190412, GW190517, and GW190521, as well as the secondary black hole intervals for GW190521 are also shown~\citep[shown as dashed contours,][]{GWTC2cat, GW190521a, GW190521b}.

By comparing the gravitational-wave detection-weighted binary black hole populations, we see a number of distinct features. 
First, the observed population is restricted to black holes with individual masses $\sim5$---$200\,M_\odot$.
These selection biases are well-known and consistent between both scenarios presented here~\citep{Messenger2012, Farr2019}. 
Additionally, the black hole mass-spin correlation is still observable in the population in the presence of ultralight bosons. 
Finally, there is an increase in the number of heavier black hole detections in the ultralight boson scenario. 
To quantify the increased number of heavier black-hole mergers populations, we compute the merger rates of both the total population, $\mathcal{R}$, as well as the intermediate mass ($100 < M/M_\odot < 1000$) population; please refer to App.~\ref{app:rates} for the details. 
We summarize the merger rates calculated and present the observed rates~\citep{GW190521b, GWTC2pop, GWTC_IMBH} in Tab.~\ref{tab:mergerrates}. 
The values calculated in the absence of ultralight bosons are consistent with the results from~\cite{Antonini2018hierarchical}.

\begin{table*}[t!]
    {\centering
    \caption{Total and intermediate mass black hole merger rates from observations of gravitational waves from stellar mass black hole mergers~\citep{GW190521b, GWTC2pop, GWTC_IMBH} compared to the calculated merger rates from our simulated population (with 68\% credible intervals). 
    The observed intermediate mass black hole merger rate is a single-event rate determined from GW190521~\citep{GW190521b, GWTC_IMBH}. 
    While the total merger rate cannot be justified by hierarchical mergers in nuclear star clusters, the observed rate of intermediate mass mergers might be explained by such events. The inclusion of superradiance due to the existence of bosons with $\mu = 4.1\times10^{-14}$\,eV leads to a doubling in the intermediate mass black hole merger rate for our simulated population. Here, $f_\textrm{nMBH}$ corresponds to the fraction of galaxies without a central massive black hole.}
    \begin{tabular}{c c c}\hline\hline
        & Total merger rate, $\mathcal{R}$ & IMBH merger rate, $\mathcal{R}_{\rm IMBH}$ \\
        & [Gpc$^{-3}$\,yr$^{-1}$] & [Gpc$^{-3}$\,yr$^{-1}$]\\
        \hline
        Observed & $23.9^{+14.3}_{-8.6}$\, & $0.08^{+0.19}_{-0.07}$ \\
        \hline
        No ultralight boson & $\approx1.2f_\textrm{nMBH}$ & $\approx0.04f_\textrm{nMBH}$\\
        $\mu = 4.1\times10^{-14}$\,eV & $\approx0.9f_\textrm{nMBH}$ & 
        $\approx0.08f_\textrm{nMBH}$\\\hline\hline
    \end{tabular}\\}
    \label{tab:mergerrates}
\end{table*}

From these calculations, nuclear star clusters, regardless of whether superradiance occurs, cannot explain the total observed merger rate from gravitational-wave detectors~\citep{GWTC2cat}. 
This result is expected, and it is anticipated that field binaries and/or dynamical mergers in globular clusters can explain the bulk of the observed binary black hole mergers~\citep{Dominik2013tma, Neijssel2019, Eldridge2019, Mapelli2020, Santoliquido2020bry,Kremer2020,Zevin2021, Wong2021, Rodriguez2021}.
However, the inclusion of superradiance leads to a doubling of the merger rate of intermediate mass black-hole mergers to $\approx0.08f_\textrm{nMBH}$\,Gpc$^{-3}$\,yr$^{-1}$ for $\mu=4.1\times10^{-14}$\,eV bosons. 
Both inferred values of $\mathcal{R}_\textrm{IMBH}$ are consistent with the single-event merger rate determined for GW190521~\citep{GW190521b, GWTC_IMBH}. 
Future gravitational-wave observations of heavy mass binary black hole systems will continue to reduce the uncertainty in the merger rate of these heavier systems. 

Finally, within our simulated population, the interpretations of gravitational-wave observations remain unchanged in the presence of ultralight bosons. 
For GW190521~\citep{GW190521a,GW190521b}, we conclude that the system is likely a 2G+2G binary black hole merger~\citep[discussed in][]{Kimball2020b, Romero-Shaw2020, Gayathri2020}.
Recently, \cite{Ng2021constraints} confirmed that the primary black holes from GW190412~\citep{GW190412} and GW190517~\citep{GWTC2cat} exclude non-self-interacting ultralight scalar bosons with masses $1.3$ --- $2.7\times10^{-13}$\,eV. From our simulations, we suspect GW190517 might be a 2G+1G merger too light to be spun-down through superradiance. 
This is tentatively supported by the results from \cite{Kimball2020b}. 
GW190412 is inconsistent with our simulated population irrespective of the presence of ultralight bosons --- likely a direct result of our mass cut-off or spin distribution used.
We can nevertheless interpret GW190412 as a direct result of hierarchical and/or dynamical formation~\citep[as in][]{GW190412, Safarzadeh2020qrc,Rodriguez2020, Zevin2020gxf, Gerosa2020}. For the most impactful boson masses from our analysis, GW190412 and GW190517 do not provide any information about their presence. 

\subsection{Detectability}

Currently, the existence of ultralight bosons is purely hypothetical. 
In the mass range we are focused on in this manuscript, $\mu\sim2\times10^{-14}$ --- $2\times10^{-13}$\,eV, their detection would likely be made by gravitational-wave detectors via either the direct observation of continuous gravitational waves, as a stochastic background, or as a feature of the population~\citep{Isi2018direct, Ng2019constraints, Ng2021constraints, Tsukada2019stochA, Sun2020, Brito2017stoch, Tsukada2021stoch}. 
Other ultralight boson searches would likely cover a different boson mass range, which impact black holes of different masses.
The detection of ultralight bosons in this energy range is likely only possible with future generations of gravitational-wave detectors~\citep{Isi2018direct, Sun2020}. 
\cite{Isi2018direct} found that the horizon distance for continuous gravitational-wave emission from boson clouds with masses $10^{-14}$ -- $10^{-13}$\,eV is $\lesssim 100$\,Mpc for aLIGO detectors~\citep{aasi2015characterization}. 
Third-generation gravitational-wave detectors such as Cosmic Explorer~\citep{Reitze2019iox} or Einstein Telescope~\citep{Maggiore2019uih} would be able to detect continuous gravitational-wave emission from the desired ultralight bosons out to a horizon of $\sim1$\,Gpc. 
Analysing the mass-spin distributions of the stellar mass black holes may be a more successful method for indirectly determining the existence of ultralight bosons~\citep{Ng2019constraints, Ng2021constraints}. 
\cite{Ng2019constraints} found that $25^{+95}_{-15}$ or $80^{+210}_{-70}$ gravitational-wave detections with signal-to-noise ratios exceeding 30 were required to determine the presence of ultralight bosons with $\mu=10^{-13}$\,eV, for two fiducial first-generation spin distributions.
It is unlikely that these detection numbers can be applied to a $\sim10^{-14}$\,eV boson detection prediction.
While current ground-based observatories provide ultralight boson detection opportunities from nearby known black holes, future gravitational-wave detectors provide a significantly greater chance. 

\section*{Acknowledgements}

We thank Richard Brito for insightful discussions about black hole superradiance. 
We also thank Susan Scott and Karl Wette for helpful discussions about ultralight boson cloud detection prospects. This work is supported through Australian Research Council (ARC) Centre of Excellence CE170100004. EP acknowledges the support of ARC CE170100004's COVID-19 support fund. PDL is supported through
ARC Future Fellowship FT160100112, and ARC Discovery Project DP180103155. KK is supported by an NSF Astronomy and Astrophysics Postdoctoral Fellowship under award AST-2001751. 

This research has made use of data, software and/or
web tools obtained from the Gravitational Wave Open
Science Center (https://www.gw-openscience.org), a service of LIGO Laboratory, the LIGO Scientific Collaboration and the Virgo Collaboration. Computing was performed on LIGO Laboratory computing cluster at California Institute of Technology. 

\appendix

\section{Cluster properties under H\'enon's principle}~\label{app:a}

\subsection{Cluster evolution}

Using H\'enon's principle, the heating rate from  binary black hole systems in the core of the cluster is balanced against the energy flow into the cluster as a whole~\citep{Henon1961, Henon1975, Gieles2011, Breen2013}. The rate of energy generation, $\dot{E}$, is therefore directly related to the properties of the cluster,
\begin{equation}
    \dot{E} = \zeta \frac{|E|}{\tau_\textrm{rh}},\label{eq:edot}
\end{equation}
where $E \simeq -0.2 G M_\textrm{cl}^2/ \rh$ is the total energy of the cluster, $M_\textrm{cl}$ is the cluster mass, $r_h$ is the half-mass radius, $\zeta \simeq 0.1$~\citep{Gieles2011,Alexander2012} is a dimensionless factor, and $\trh$ is the relaxation timescale of the cluster which can be approximated as~\citep{Spitzer1971, Antonini2018hierarchical}
\begin{equation}
    \tau_\textrm{rh} = 0.138\sqrt{\frac{M_\textrm{cl}\rh^3}{G}}\frac{1}{\langle{m_*}\rangle \psi \ln\Lambda}. \label{eq:trh}
\end{equation}
Here, $\langle{m}_*\rangle$ is the mean stellar and compact object mass within the cluster and $\ln\Lambda \simeq 10$ is the Coulomb logarithm. 
The factor $\psi$ is related to the distribution of masses within the half-mass radius and approximated in \cite{Spitzer1971} as $\psi = \langle{m^{2.5}_*}\rangle/\langle{m_*}\rangle^{2.5}$.
For the remainder of the manuscript, we follow~\cite{Antonini2018hierarchical} and assume low-mass stars dominate the mass distribution of the cluster, such that $\langle{m_*}\rangle = 0.6\,M_\odot$, and $\psi \simeq 5$. 

Finally, by defining the half-mass density, $\ph = 3\mcl/8\pi \rh^3$, and escape velocity, $v_\textrm{esc} \propto \sqrt{\mcl/\rh}$, we can express the initial heating rate, relaxation time, and escape velocity at $t=t_0$ uniquely in terms of the initial cluster density and its mass. From Eqs.~\eqref{eq:edot} and~\eqref{eq:trh}~\citep{Antonini2018hierarchical}, 
\begin{subequations}~\label{eq:initial}
\begin{align}
    &\dot{E}_0 \simeq 2.3\times10^5\,M_\odot\,(\textrm{km s}^{-1})^2\,\textrm{Myr}^{-1}M_5^{2/3}\rho_{5,0}^{5/6},\\
    &\tau_{\textrm{rh},0} \simeq 7.5\,\textrm{Myr}\,M_5\rho_{5,0}^{-1/2},\\
   &v_{\textrm{esc},0} \simeq 50\,\textrm{km s}^{-1}\,M_5^{1/3}\rho_{5,0}^{1/6},
\end{align}
\end{subequations}
where $M_5 \equiv \mcl/10^5\,M_\odot$ and $\rho_{5,0} \equiv \rho_{\textrm{h},0}/10^5\,M_\odot\,\textrm{pc}^{-3}$. Here, $t_0$ corresponds to the time at which the first black hole binaries begin to heat the cluster (i.e. when they have entered the cluster's core)~\citep{Breen2013}. 
Furthermore, we have implicitly assumed that mass loss of the cluster is a negligible contribution to the evolution of the cluster. Below, we explicitly require a constant $\mcl$ in the calculation of the cluster model's evolution. 
The proportionality constant for the initial escape velocity is derived from the~\cite{King1966} cluster model with $W_0 = 7$, where $W_0$ is the central value of the dimensionless form of the cluster's gravitational potential, uniquely describing the potential's shape. This model is used throughout for the calculation of the dynamical features of the clusters. 

To derive the time dependence of the cluster properties, we assume no mass loss (constant $\mcl$) and that the cluster is always in virial equilibrium. Under these assumptions, the rate of expansion of the cluster is related to the heating rate as~\citep{Henon1965}
\begin{equation}
    \frac{\dot{r}_\textrm{h}}{r_\textrm{h}} = \frac{\dot{E}}{|E|} = \frac{\zeta}{\trh},~\label{eq:de}
\end{equation}
following Eq.~\eqref{eq:edot}.
Fixing the cluster mass, we have $\trh \propto r_h^{3/2}$, and solve Eq.~\eqref{eq:de} for the time evolution of the half-mass radius from $t_0$, 
\begin{equation}
    \rh(t) = r_{\textrm{h},0} \Big(\frac{3}{2}\frac{\zeta(t-t_0)}{\tau_{\textrm{rh},0}} + 1\Big)^{2/3},\label{eq:rht}
\end{equation}
when $t > t_0$. 
When $t < t_0$, the cluster has not yet extracted energy from the binary population and is assumed to have the initial half-mass radius. 
Eq.~\eqref{eq:rht} can be directly substituted into the expressions for the heating rate and escape velocity to find
\begin{equation}
    \dot{E}(t) = \dot{E}_0 \Big(\frac{3}{2}\frac{\zeta(t-t_0)}{\tau_{\textrm{rh},0}} + 1\Big)^{-5/3},\label{eq:edott}
\end{equation}
and
\begin{equation}
    v_{\textrm{esc}}(t) = v_{\textrm{esc},0} \Big(\frac{3}{2}\frac{\zeta(t-t_0)}{\tau_{\textrm{rh},0}} + 1\Big)^{-1/3}.\label{eq:vesct}
\end{equation}

\subsection{Binary black hole evolution}

After a binary black hole system has formed, the system undergoes dynamical hardening~\citep{Leigh2018}. 
This process tightens the binary orbit while heating the cluster via H\'enon's principle~\citep{Henon1975}. 
This implies that the rate of energy loss from the binary is approximately equal to the heating rate of the cluster in Eq.~\eqref{eq:edot},
\begin{equation}
    \dot{E}(t) \simeq -\dot{E}_\textrm{bin}.\label{eq:energyeq}
\end{equation}
The total energy of the binary during the process of dynamical hardening is 
\begin{equation}
    E_{\textrm{bin}} = - \frac{GM_1M_2}{2a},\label{eq:Ebin}
\end{equation}
where $a$ is the semimajor axis of the binary's orbit. 
Therefore, differentiating both sides of Eq.~\eqref{eq:Ebin} and applying Eq. \eqref{eq:energyeq} to express the semimajor axis evolution of the binary as a function of the heating rate of the cluster leads to~\citep{Antonini2018hierarchical}
\begin{equation}
    \dot{a} = -\frac{2a^2}{GM_1M_2}\dot{E}(t).\label{eq:dota}
\end{equation}
We integrate this expression with respect to the semimajor axis from the initial semimajor axis of formation, $a_h$, to the minimum semimajor axis for which dynamical hardening dominates, $a_m$. 
The heating rate is assumed to be approximately constant over the binary's evolution. 
This defines the timescale of dynamical hardening for a binary to be
\begin{equation}
    \tau_\textrm{dyn} = -\int^{a_h}_{a_m} \frac{GM_1M_2}{2a^2}\dot{E}^{-1}(t)\,\dd a \simeq \frac{GM_1M_2}{2a_m}\dot{E}^{-1}(t), 
\end{equation}
where we have used the fact that $a_h \gg a_m$, as seen in Eq.~\eqref{eq:tdyn}. 

The lower bound semimajor axis, $a_m$, is given by
\begin{equation}
    a_m = \max(a_{\textrm{ej}}, a_{\textrm{GW}}),
\end{equation}
where $a_\textrm{ej}$ is the semimajor axis for which the binary system is ejected from the cluster through binary-single interactions, and $a_\textrm{GW}$ is the semimajor axis at which the energy loss from dynamical hardening is equal to the energy loss from gravitational-wave emission. 
If the semimajor axis of the binary decreases below $a_{\textrm{ej}}$, then the binary is ejected from the cluster prior to merging~\citep{Quinlan1996, Miller2002, Miller2009}. The value of $a_\textrm{ej}$ is given by~\citep{Antonini2016Nscs}
\begin{equation}
    a_\textrm{ej} \simeq 0.2 \frac{M_1M_2}{M_{123}}\frac{q_3}{v_\textrm{esc}^2(t)},
\end{equation}
where $M_{123} = M_1 + M_2 + \langle M \rangle_\textrm{core}$ is the average total mass of the binary-single interaction, and $q_3 = \langle M \rangle_\textrm{core}/(M_1+M_2)$ is the mass ratio of the interaction. 
Since we do not track the individual trajectories and interactions within the cluster, we use the average black hole mass in the core (excluding the binary), $\langle M \rangle_\textrm{core}$, to calculate $a_\textrm{ej}$. 

If $a_\textrm{GW} > a_\textrm{ej}$, the binary will merge within the cluster. 
Due to the massive and dense clusters considered in this manuscript, this occurs for almost all binaries. 
The evolution of the semimajor axis through gravitational-wave emission is given by~\citep{Peters1964}
\begin{equation}
    \dot{a}_\textrm{GW} = -\frac{65}{4}\frac{G^3M_1M_2(M_1+M_2)}{c^5a^3(1-e^2)^{7/2}}g(e),\label{eq:dotagw}
\end{equation}
where, $e$ is the eccentricity of the binary, and  
\begin{equation}
    g(e) = 1 + \frac{73}{24}e^2 + \frac{37}{96}e^4. 
\end{equation}
By equating Eqs.~\eqref{eq:dota} and~\eqref{eq:dotagw}, solving for $a$ leads to the expression~\citep{Antonini2018hierarchical}
\begin{equation}
    a_\textrm{GW}^5 = \frac{32}{5}\frac{G^4M^2_1M^2_2(M_1+M_2)g(e)}{c^5(1-e^2)^{7/2}}\dot{E}^{-1}(t).
\end{equation}
For every binary formed, we generate an eccentricity from the thermal eccentricity distribution $\propto e$~\citep{Jeans1919}. 
The timescale for gravitational-wave emission from the binary is then given by $\tau_\textrm{GW} = a_m/|\dot{a}_\textrm{GW}|$. 
Just prior to merger, the effects of superradiance on the individual black holes are incorporated. 

\subsection{Interloper ejection}~\label{app:interloper}

Although it is very unlikely that the binary system is ejected from binary-single interactions, this is not the case for interlopers. 
To account for this, the expected number of ejected interlopers as~\citep{Antonini2018hierarchical}
\begin{equation}
    N_{\textrm{ej}} \approx 6\,\ln(\frac{a_\textrm{ej}}{q_3^2a_m}),
\end{equation}
where the interloper mass ratio, $q_3$, is calculated from the mean mass of black holes in the core. 
With the approximate number of ejected black hole interlopers calculated, we discard $N_\textrm{ej}$ black holes randomly. 
Generally, this process would favor the ejection of low mass black holes. 
However, given that the majority of black holes in the cluster are from the first generation, this averaged procedure does not change the expected results. 
The key difference with the inclusion of interloper ejection is a restriction on the upper mass of the black hole formed through hierarchical black hole mergers. 
Since this process removes a fraction of black holes from the cluster, in some simulations this limits the available black holes for hierarchical growth. 

\section{recoil velocity calculation}\label{app:kick}

We utilize the {\tt \sc Precession} code~\citep{Gerosa2016} to determine the final recoil velocity of the remnant black hole.
The kick velocity fitting formula are constructed from mass weighted spin vector combinations~\citep{Gerosa2016},
\begin{align}
\bm{\Delta} &= \frac{\chi_1\bm{\hat{S}}_1 - q\chi_2\bm{\hat{S}}_2}{1+q},\\
\bm{\tilde{\chi}} &= \frac{\chi_1\bm{\hat{S}}_1 + q^2\chi_2\bm{\hat{S}}_2}{(1+q)^2},
\end{align}
where $\chi_{1,2}$ are the spin magnitudes of the individual black holes, $\bm{\hat{S}}_{1,2}$ are their associated unit vectors, and $q = M_2/M_1$ is the mass ratio. These vectors can also be decomposed into the components parallel and orthogonal to the orbital angular momentum, $\bm{L}$, given as $\tilde{\chi}_\| = \bm{\tilde{\chi}}\cdot\bm{\hat{L}}$, $\tilde{\chi}_\perp = |\bm{\tilde{\chi}}\times\bm{\hat{L}}|$, $\Delta_\| = \bm{\Delta}\cdot\bm{\hat{L}}$, and $\Delta_\perp = |\bm{\Delta}\times\bm{\hat{L}}|$. 

The formulation of the final magnitude of the kick velocity of the black hole, $v_k$, is constructed from the recoil velocities from mass, $v_m$, and spin, $v_{s\|}$ and $v_{s\perp}$, asymmetries in the binary system. 
The velocity component from mass asymmetry lies perpendicular to the orbital angular momentum vector, whereas the spin component has some contribution both parallel to the vector, and in the orbital plane of the binary. This leads to the simple expression for the magnitude of the kick velocity~\citep{Campanelli2007kick}, 
\begin{equation}
    v_k = \sqrt{v_m^2 + 2v_mv_{s\perp}\cos\beta + v_{s\perp}^2 + v_{s\|}^2},~\label{eq:vk}
\end{equation}
where $\beta$ is the angle between the components from mass and spin asymmetry orthogonal to the orbital angular momentum. 
Each of the individual terms in Eq.~\eqref{eq:vk} has been determined through analytical fits to numerical relativity simulations. 
The velocity components $v_m$, $v_{s\|}$, and $v_{s\perp}$ are calculated as
\begin{align}
    v_m &= A\eta^2\frac{1-q}{1+q}(1+B\eta),\\
    v_{s\perp} &= H\eta^2\Delta_\|,\\
    v_{s\|} &= 16\eta^2[\Delta_\perp(V_{11} + 2V_A\tilde{\chi}_\| + 4V_B\tilde{\chi}_\|^2 + 8V_C\tilde{\chi}_\|^3) \nonumber\\
    &\,+ 2\tilde{\chi}_\perp \Delta_\|(C_2+2C_3\tilde{\chi}_\|)]\cos\Theta,\label{eq:vsparallel}
\end{align}
where $\eta = M_1M_2/(M_1+M_2)^2$, and the coefficients have been found through fitting to numerical relativity simulations: $A = 1.2\times10^4$\,km\,s$^{-1}$, $B = -0.93$~\citep{Gonzalez2007kick}, $H = 6.9\times10^3$\,km\,s$^{-1}$~\citep{Lousto2008}, $V_{11} = 3677.76$\,km\,s$^{-1}$, $V_A = 2481.21$\,km\,s$^{-1}$, $V_B = 1792.45$\,km\,s$^{-1}$, $V_C = 1506.52$\,km\,s$^{-1}$\citep{Lousto2012a}, $C_2 = 1140$\,km\,s$^{-1}$, $C_3 = 2481$\,km\,s$^{-1}$~\citep{Lousto2012kick}, and $\beta = 145^\circ$~\citep{Lousto2008}. 
Finally, $\Theta$ is the angle between $\bm{\Delta}\times\bm{\hat{L}}$ and the infall direction of the black holes at merger, with an additional offset of $\sim200^\circ$~\citep{Brugmann2008, Lousto2009}. This angle is assumed to be uniformly distributed from $0$ to $\pi$~\citep{Gerosa2016}, and is therefore drawn at random for each merger calculation. 

\section{Calculating the fraction of hierarchical-growth supported clusters}~\label{app:fnc}

We use Monte Carlo integration to determine the fraction of the individual nuclear star cluster property distributions, $p_i(R_\textrm{eff}, M_\textrm{cl})$, contained within the contour (which itself has uncertainty), where $i$ indexes the cluster. 
To estimate the uncertainty on the contours, the uncertainty on the fraction of simulations that demonstrate hierarchical growth must first be calculated. 
By assuming the fraction follows a binomial distribution, the uncertainty on $f_\textrm{hier}$ is estimated with a Wilson score interval~\citep{wilson1927probable}.
The median and confidence interval are fit with a generalized extreme value (GEV) distribution following~\cite{possolo2019asymmetrical}. 
This allows us to draw $1000$ possible fractions at each point in the parameter space, from which we create $1000$ possible hierarchical growth contours. 
We fit the $N_\textrm{NSC} = 228$ nuclear star cluster properties~\citep{Georgiev2016} with GEV distributions~\citep{possolo2019asymmetrical} as an approximation for $p_i(R_\textrm{eff}, M_\textrm{cl})$ to generate $N_p=4\times10^3$ plausible $(R_\textrm{eff}, M_\textrm{cl})$ values for each cluster. 
We compute the integral as
\begin{align}
    f_{\textrm{NSC},j}&\equiv \frac{1}{N_\textrm{NSC}}\sum_{i=1}^{N_\textrm{NSC}}\int p_i(R_\textrm{eff}, M_\textrm{cl}) \Theta_j(R_\textrm{eff}, M_\textrm{cl})\,\dd R_\textrm{eff} \,\dd M_\textrm{cl}\nonumber\\
    f_{\textrm{NSC},j}&\approx \frac{1}{N_\textrm{NSC}N_p}\sum_{i=1}^{N_\textrm{NSC}}\sum_{k=1}^{N_\textrm{p}}\Theta_j(R^{i,k}_\textrm{eff}, M^{i,k}_\textrm{cl}),\label{eq:frac}
\end{align}
where $\Theta_j(R^{i,k}_\textrm{eff}, M^{i,k}_\textrm{cl})$ evaluates to one for a point within the $j^\textrm{th}$, $f_\textrm{hier} > 50\%$ contour, and zero otherwise. The value of $f_\textrm{NSC}$ is taken as the median from the distribution $f_{\textrm{NSC},j}$, with an uncertainty set by the distribution.

\section{Calculating merger rates}\label{app:rates}

From the synthetic population presented in Sec.~\ref{subs:presentpop}, the merger rate is estimated following \cite{Antonini2018hierarchical} as 
\begin{equation}
    \mathcal{R} \approx \dv{N_\textrm{gx}}{V_c} \frac{f_\textrm{nMBH}}{N_\textrm{NSC}}\sum^{N_\textrm{NSC}}_{i=1}\frac{1}{\tau_{\textrm{dyn},i}},\label{eq:r1}
\end{equation}
where $\dd N_\textrm{gx}/\,\dd V_c\approx0.01\,$Mpc$^{-3}$~\citep{Conselice2005} is the number density of nucleated galaxies, and $f_\textrm{nMBH}$ is the fraction of galaxies without a central massive black hole. For each nuclear star cluster, we compute the reciprocal of the binary formation timescale, $\tau_\textrm{dyn}$, as an approximate for the rate of binary mergers within the cluster, since a single binary dominates the cluster evolution at any one time. 
Here, $f_\textrm{nMBH}$ is kept as a parameter in the expression as its value is somewhat unconstrained, though expected to be $\sim0.2$ --- 1~\citep{Seth2008, Antonini2015, Nguyen2018}. 
Furthermore, the merger rate for which at least one member of the binary is within the black hole mass range $10^2\,M_\odot < M < 10^3\,M_\odot$ is calculated as
\begin{equation}
    \mathcal{R}_\textrm{IMBH} \approx \dv{N_\textrm{gx}}{V_c} \frac{f_\textrm{nMBH}}{N_\textrm{NSC}}\sum^{N_\textrm{NSC}}_{i=1}\frac{\Theta(10^2 < M_{1,i}/M_\odot < 10^3)}{\tau_{\textrm{dyn},i}},\label{eq:r2}
\end{equation}
where $\Theta(10^2 < M_{1,i}/M_\odot < 10^3)$ evaluates to one when the condition is satisfied and zero otherwise. 
This mass-range corresponds to the lower end of the intermediate mass black hole (IMBH) range \citep[e.g.,][]{Greene2020}. 
The merger remnant (and possibly also the primary black hole) for GW190521 lies conclusively within this range~\citep{GW190521a, GW190521b}. 
The calculation in Eq.~\ref{eq:r2} can be compared to the observed IMBH merger rates.

\bibliography{refs}

\end{document}